\documentclass[a4paper,12pt]{article}
\usepackage{graphicx}
\usepackage{amssymb}
\usepackage{a4}

\title{The quantum Neumann model: asymptotic analysis.}
\date{July 2005}
\author{Marc Bellon and Michel Talon\thanks{LPTHE, CNRS et Universit\'es
Paris VI--Paris VII (UMR 7589),
 Bo\^{\i}te 126,
 4 place Jussieu, F-75252 PARIS CEDEX 05}
}

\begin{document}

\maketitle

\begin{abstract}
We use semi--classical and perturbation methods to establish the quantum
theory of the Neumann model, and explain the features observed in previous
numerical computations.
\end{abstract}
\vfill
LPTHE--05--17
\eject

\section{Introduction}

In a previous publication~\cite{BT05}, we have explained how to determine
numerically the spectrum of the quantum Neumann model. The Neumann model
describes a particle on a sphere subject to a quadratic potential. It has been
introduced in the nineteenth century by Neumann~\cite{Neumann}, who showed that
it is integrable.
More recently, Moser~\cite{Moser} and Mumford~\cite{Mu84} have shown how it fits
in the modern approaches to algebraically integrable systems and generalized the model
to any number of dimensions.  More to our point, the quantum version of the
model is also integrable~\cite{AvTa90,BaTa92}. This means that
its solution reduces to the resolution of a one dimensional Schr\"odinger equation,
as it is the generic behaviour for algebraically integrable systems. The notable
feature compared to the
textbook examples is that the different conserved quantities, which all
appear in the same equation, must be solved for simultaneously. This is a non
trivial problem
which has been attacked in~\cite{Gurarie}. We will precise and prove some
statements made in this paper. Moreover, the Neumann model is related with
St\"ackel systems as noted by Gurarie and fully developed recently
in~\cite{DuVa05}.

We will start with a brief presentation of the model and the equations which
have to be solved. The physics of the model is driven by a general scale
parameter~$v$, which is proportional both to the strength of the interaction
and the radius of the sphere. Small values of $v$ correspond to a quantum
regime and are next studied by perturbation theory. Then, we introduce
the semi--classical analysis which is appropriate for large values of $v$. The 
comparison of energy levels obtained by semi--classical methods and by the
numerical solution of the Schr\"odinger equation shows that it is indeed the
case. However, we find a small constant discrepancy between the two. In a last
part, we recover this constant by analytic studies of the model at large $v$
according to these two methods.

\section{The model.}

The Hamiltonian of the system is given in terms of angular momenta
$J_{kl} = x_k \dot{x}_l - x_l\dot{x}_k$:
\begin{equation}
H = {1 \over 4} \sum_{k \neq l} J_{kl}^2 ~+ ~ 
{1 \over 2} \sum_k a_k x_k^2
\label{hamiltonian}
\end{equation}
with the constraint $\sum x_k^2=r^2$. We assume $a_1<a_2<\cdots<a_N$. One can
always shift all the $a_k$ by a
constant since this merely adds a constant to $H$. The system is classically
and quantum mechanically integrable since the following quantities commute,
classically and quantically:
\begin{equation}
F_k = x_k^2 + \sum_{l \neq k} {J_{kl}^2 \over a_k - a_l}, \quad H = {1\over2}
\sum a_k F_k
\label{conserved}
\end{equation}

To solve the equations of motions, one introduces the ``separated" variables
$t_k$ given
as the roots of:
$$ \sum_k {x_k^2 \over t-a_k}=0$$
Positivity of $x_k^2$ implies that there is exactly one root in each interval
$[a_k,a_{k+1}]$.
It has been shown in~\cite{BaTa92} that  finding the common eigenvector 
for the $F_k$ with eigenvalue $f_k$ reduces to solving the unique
one--dimensional Schr\"odinger equation:
\begin{equation}
\left[{d^2\over dt^2}+{1\over 2}\sum_k{1\over t-a_k}\;{d\over dt}
-{1\over 4\hbar^2}\sum_k{f_k\over t-a_k}\right]\,\Psi(t)=0
\label{lame}
\end{equation}
This is a linear differential equation with singularities at the $a_k$ and at
infinity. The situation is explained in more details in~\cite{BT05}.

It can be useful to write the potential term in the form:
\begin{equation} \label{produit}
\sum _k{f_k\over t-a_k}= v\; {\prod (t-b_l)\over \prod (t-a_k)}
\end{equation}
with $v = \sum_k f_k = \sum_k x_k^2=r^2$.
Classically the action $S$ satisfies the separated
Hamilton-Jacobi equation:
$$ {1\over 2}\left({dS \over dt}\right)^2 = - v\; {\prod (t-b_l)\over \prod
(t-a_k)}$$
In Figure~1 we draw the different possibilities for the potential in
the Neumann case $N=3$, with oscillator strengths 0, 1 and $y$.

\begin{figure}[t]
\includegraphics[width=12cm]{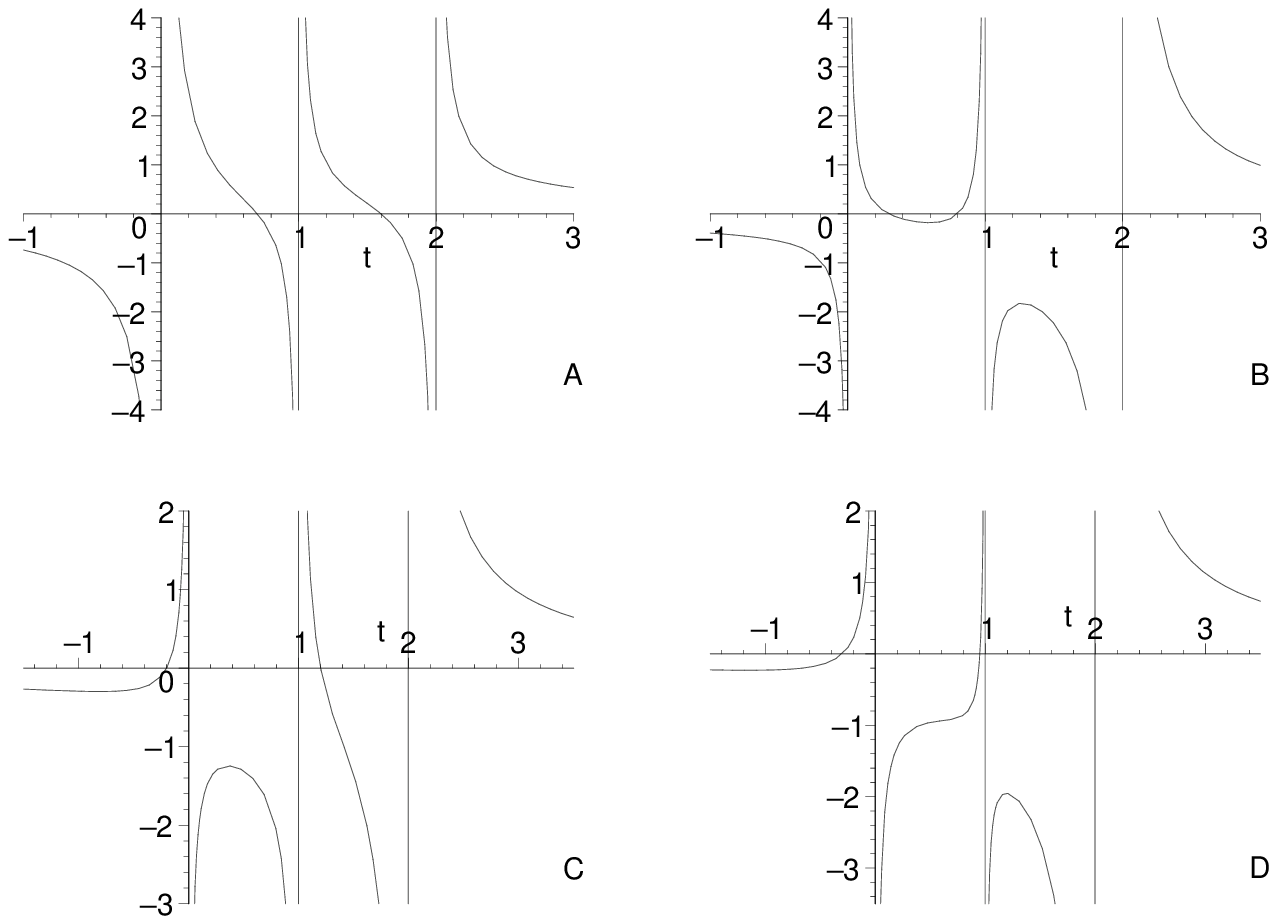}
\label{figs}\caption{The four different possibilities for $t_1,t_2$, with
$y=2$.}
\end{figure}

For $v$ large, we are in case A), $b_1$ is close to 1 and $b_2$ is close to
$y$. Classically the allowed region is such that for each $t_k$ one has
$P(t)=\prod (t -a_n)\prod(t-b_m) <0$ so that, since $t_k\in [a_k,a_{k+1}]$ 
one has $b_1\leq t_1 \leq 1$ and $b_2 \leq t_2 \leq y$. As $v$ is lowered, both
$b_1$ and $b_2$ move to the left, and when $v\to 0$, one has $b_1 \to
-\infty$ so that one is in case C) or D). One goes from case D) to case
A) either through the cases B) or C). In case B), $b_1$ and $b_2$ both belong
to the interval $[0,1]$ so that the allowed region corresponds to $b_1\leq t_1
\leq b_2$ and $t_2 $ describes the whole interval $[1,y]$. In case C) and D),
$b_1$ is negative and $b_2$ is on either side of 1. One of the $t_k$ describes
the whole interval and the other is constrained away from 1 by $b_2$.
It is easy to check that no other 
configuration is compatible with the existence of $t_1$ and $t_2$ satisfying
$P(t)<0$ in the appropriate intervals. In the general case, we will show in the
sequel that all the $b_i$ are real and belong to the interval
$[a_{i-1},a_{i+1}]$, with $a_0$
supposed to be $-\infty$. There is therefore a total of $2^{N-1}$ possibilities
for the relative positions of the $b_i$ and $a_k$.

At $v=0$, the system reduces to a free particle on the sphere. The energy
levels are given by the usual formula ${1\over 2}j(j+1)$ for spherical
harmonics, but with a different
choice of basis in the $(2j+1)$-dimensional eigenspace. These basic
functions are called spheroidal harmonics. These harmonics diagonalize the
perturbation by the potential, so that energy levels separate linearly for
small $v$.

For large $v$, the potential term becomes dominant and the point is restricted
to the vicinity of the two opposite poles $x_1=\pm r$, $x_2=0$, $x_3=0$. In the
next approximation, we can neglect the curvature of the sphere and we have two
independent harmonic oscillators in $x_2$ and $x_3$. The configuration around
the two poles are linked by an exponentially small tunnelling amplitude, so that
all states come in almost degenerate pairs.

\section{Small $v$ perturbation theory.}

In the case $v=0$, the singularity at infinity of the equation~(\ref{lame})
becomes regular. Even if the behaviour of the solution at infinity has no
direct bearing on the physical wave function, it appears important for the
characterization of the solutions and provides the quantization condition for
the energy.

Indeed, recall that around a regular singularity at $x=0$, a second order
differential equation has two independent solutions of the form $x^\alpha(1+ b x
+ \cdots)$, with the exponents $\alpha$ roots of a second order equation.
At the finite singularities $a_k$, the exponents are 0 and $1/2$ with
respective monodromies $+1$ and $-1$. In our previous work~\cite{BT05}, we
determined that
to produce a single valued wave function on the sphere, the solutions must be
of definite monodromy at each of the points $a_k$. Monodromy at infinity is the
product of the monodromies at the $a_k$ and is therefore $\pm 1$. Since the
exponent at infinity is solution of:
\begin{equation}
	\alpha ^2 + \alpha ({N\over 2}-1) - { E \over 2 \hbar^2}=0
\end{equation} 
and $\alpha$ must be integer or half integer to get the right monodromy, we
conclude that the energy $E$ is ${1\over2}\hbar^2 j(j+N-2)$ with any non
negative integer $j=2\alpha$. 

For a given energy $E$, there is a second root $\alpha$ which yields a
solution vanishing at infinity. However it cannot be the solution we are
seeking with diagonal monodromy at the $a_k$, since such a solution would be
single valued and regular on a compact Riemann surface. In the case where all
monodromies are one, the solution has just one singularity, a pole at infinity
and is therefore a polynomial of degree $\alpha$. These functions are known as
spheroidal harmonics of first species~\cite{WW}.

In order to find these polynomial solutions, we write them in the form $\prod
(t-\lambda_j)$. Inserting in equation~(\ref{lame}), we get:
\begin{equation}
\sum_{i\neq j} {1\over(t-\lambda_i)(t-\lambda_j)}+{\textstyle 1 \over
2}\sum_{k,j} {1\over(t-a_k)(t-\lambda_j)} - {1\over
4\hbar^2}\sum_k{f_k\over t-a_k} =0
\end{equation}
As it should be, this equation requires that $\sum_k f_k=0$ and we recover the
value for the energy from the dominant term at $t \to \infty$. This equation is
equivalent to the condition that its residues at the $a_k$ and $\lambda_j$ all
vanish. The residues at the $\lambda_j$ give a system of non-linear equations
which determine the $\lambda_j$:
\begin{equation}
\sum_k{1\over \lambda_j-a_k}+4 \sum_{h\neq j}{1\over \lambda_j- \lambda_h} = 0
\label{eqzero}
\end{equation}
The residues at the $a_k$ then determine the conserved quantities as:
\begin{equation}
f_k = 2 \hbar^2 \sum_j{1\over a_k - \lambda_j}
\end{equation}
The system of equations~(\ref{eqzero}) can be seen as the condition of
equilibrium of a system of points with logarithmic repulsive potentials, some
of which being fixed. In this formulation, it is therefore clear that there is
a unique solution with a definite number of $\lambda_j$ in each of the intervals
$[a_k,a_{k+1}]$. In the case $N=3$, this gives $d+1$ different solutions of
degree $d$. Similar calculations can be done for solutions which are of the
form $\prod_{k\in I}\sqrt{t-a_k}\prod_j(t-\lambda_j)$ where $I$ is a subset of
the singularities. This gives spheroidal harmonics of the second, third and
fourth species~\cite{WW} for $N=3$. The total number of these harmonics with
energy ${1\over2}j(j+1)$ is exactly $2j+1$, the known degeneracy of this energy
level.

Numerical solutions of the system of equations~(\ref{eqzero}) are easy to get
by a multidimensional Newton method and have been used to produce initial
solutions in our previous paper.

From the knowledge of this $v=0$ solutions, we will describe the behaviour of
the solutions for small $v$. As soon as $v$ is different from zero, the
solution gets an essential singularity at infinity. We therefore multiply the
spheroidal harmonic by the simplest function singular at infinity, namely
$\exp(pt)$. In the vicinity of $v=0$, we expect $p$ to be small and neglect
higher powers of $p$. Inserting in equation~(\ref{lame}) gives:
\begin{eqnarray*}
\sum_{i\neq j} {1\over(t-\lambda_i)(t-\lambda_j)}
+{\textstyle 1 \over2}\sum_{k,j} {1\over(t-a_k)(t-\lambda_j)} 
- {1\over4\hbar^2}\sum_k{f_k\over t-a_k} &&\\+ 2 p \sum_j{1 \over t-\lambda_j}
+{1\over 2}p\sum_k{1\over t - a_k}&=&0
\end{eqnarray*} 
The term proportional to $1/t$ in this equation immediately gives that $p$ is
proportional to $v$, so that the hypothesis that $p$ is small is valid. 
\[ 2np + {1\over2}N p -{1\over4\hbar^2}v =0
\]
Here $n$ is the degree of the polynomial. The roots $\lambda_j$ of the wave
function satisfy equations which are small perturbations of
equation~(\ref{eqzero}), hence are close to their values for
$v=0$. The term in $1/t^2$ then gives the perturbation of the energy:
\[ E =  \hbar^2 n(2n+N-2) + v { \sum a_k + 4 \sum \lambda_j \over 2(N + 4n)} \]
Similar calculations can be done for the spheroidal harmonics of the other
species and this nicely reproduces the slopes of the numerical results at small
$v$.

\section{Semiclassical analysis.}

As an integrable system, the Neumann model lends itself to a simple
semiclassical analysis writing basically that the Bohr--Sommerfeld conditions
quantize the Liouville tori. Equivalently, the separated Schr\"odinger equation
can be analyzed by the WKB method. 

\subsection{The WKB treatment.}

The presence of a singular factor in the
first derivative term of the equation~(\ref{lame}) seems to be a problem for
the application of the WKB method in our case, but we will show that in
contrast to what happens in the spherical symmetry case~\cite{Lan38}, this does
not cause troubles. In a first approach, we write an expansion in powers of
$\hbar$ for the action $S$ such that $\Psi = \exp (iS/\hbar)$:
\begin{equation}
\label{expans} S = S_0 + \hbar S_1 + \hbar^2 S_2 + \cdots
\end{equation} 
As usual the dominant term in $\hbar^{-2}$ does not depend on the first
derivative term in the equation and gives for $S_0$:
\[ S_0 = \int dt \;\sqrt{-v\prod(t-b_i) \over 4 \prod (t-a_k)} \]
The following term gives:
\[ i S_0'' - 2 S_0'S_1' + {1\over2} i S_1' \sum{1\over t- a_k} = 0 \]
Due to the form of $S_0$, the singular terms at $a_k$ cancel and one gets:
\[ \exp( iS_1) =  \prod (t-b_j) ^{-{1\over4}} \]
At this order, the wave function is regular at the $a_k$ and one can also show
that the following correction to $S$ is regular at $a_k$. In contrast, the
first correction is mildly singular at the turning point $b_l$ and the
following correction $S_2$ has so bad singularities that the expansion breaks
down. In the cases where the $b_l$ are away from the $a_k$, the usual
connection formulae around the turning point~\cite{Lan38} justify an additional
$\pi/4$
phase factor in the quantization conditions.

Another approach is to desingularize the equation~(\ref{lame}) by introducing a
new variable $x$ defined by the (hyper-)elliptic integral~\cite{WW}:
\begin{equation}\label{ellparam}
x = \int^t {dt \over \sqrt{\prod(t-a_k)}}
\end{equation}
 The application of the chain rule gives:
\[ {d^2\Psi\over dt^2}+{1\over 2}\sum_k{1\over t-a_k}\;{d\Psi\over dt} =
{1\over\prod(t-a_k)} {d^2\Psi\over dx^2} \]
so that the Schr\"odinger equation becomes:
\begin{equation}\label{ellequ}
{d^2\Psi\over dx^2} - {v\over 4\hbar^2} \prod ( t-b_l) \Psi =0
\end{equation}  
This equation has neither singularities nor first derivative term, so that the
usual WKB formula gives:
\[ \Psi ( x) \simeq \left( \prod (t-b_j) \right) ^{-{1\over4}} \exp\left(
{1\over\hbar} \int dx \sqrt{{v\over 4}\prod(t-b_i) }\right) \]
It should be emphasized that the variation of $x$ is alternatively real and
purely imaginary when $t$ crosses the $a_k$, introducing additional $i$
factors in the action. In the traditional $N=3$ case, $x$ is an elliptic
integral and $t$ can be expressed with Weierstra\ss \ elliptic functions. When
$t$ goes from $-\infty$ to $+\infty$, $x$ describes the boundary of the
half--periods rectangle. In the cases where there are more singularities, it is
possible that the $t$ variable is not well defined as an inverse function of
$x$, however this is of no consequence for the study of the Schr\"odinger equation
which proceeds separately on each of the intervals $[a_k,a_{k+1}]$.

\subsection{Remarks on the zeroes of the potential.}

The above elliptic formulation allows to demonstrate quickly the claim we 
have made that there must be in each interval $[a_j,a_{j+1}]$ a region
where the product $P(t)=\prod (t -a_n)\prod(t-b_m)$ takes negative values.
Recall from~\cite{BT05} that the wave function is either even or odd under
each parity operation $x_k \to -x_k$. In terms of the $t$ variable, this
translates into monodromy around $a_k$. Now, the $x$ variable of the elliptic
parametrization behaves as $x(a_k)+c_k\sqrt{t-a_k}$ in the vicinity of $a_k$ so
that the wave function is even or odd as a function of $x-x(a_k)$, which means
that either $\Psi$ or its derivative vanishes at each of the $x(a_k)$.

Multiplying eq.~(\ref{ellequ}) by $\Psi$ and integrating
by parts, we then get for each $k$: 
\begin{equation} \label{ene}
\int_{x(a_k)}^{x(a_{k+1})} [\Psi'^2(x) + {v\over 4\hbar^2}\prod(t(x)-b_l)
\Psi^2(x)]
dx = 0
\end{equation}
Under the elliptic parametrization, the image of the real axis is a sequence
of alternatively real and purely imaginary intervals, according to the sign
of the product $\prod (t-a_k)$. The $\Psi'^2(x)$ term has therefore the same
sign as this product. Since $\Psi$ is real, Eq.~(\ref{ene}) cannot be satisfied if the
potential remains of the same sign as the kinetic term.   This is
equivalent to the condition $P(t)<0$ at some point of the interval
$[a_j,a_{j+1}]$.
Finally, we prove that this allows to establish that  quantum mechanically the
$b_j$ obey the constraints we have discussed in the framework of the classical analysis.

The relative position of the $a_k$ and $b_j$ depends on the signs of the $f_k$.
The only one with a fixed sign is $f_N$, which is positive. All combinations of
the signs of the other ones are possible, which gives $2^{N-1}$ possibilities.
We will show that each of these sign combinations corresponds to a unique set
of relative positions of the $b_j$. The fact that we identify $N-1$ necessary
real zeroes of the potential shows that all $b_j$ must be real.

More precisely, we will show that $b_j$ belongs to the interval $]a_j,a_{j+1}[$
if $f_j$ is positive and to the interval $]a_{j-1},a_{j}[$ otherwise. This
means that we have to show that in the interval $]a_j,a_{j+1}[$, there is
exactly one zero if $f_j$ and $f_{j+1}$ are of the same sign and two if $f_j$
is positive and $f_{j+1}$ is negative. If the $f$ are of the same sign, the
potential reaches opposite infinite limits at the ends of the interval, so that
it must have a zero in between. In the other case, the potential goes to
positive infinity at the two ends of the interval and we proved that the
potential must be negative somewhere in the interval. This is only possible if
the potential has two zeroes which we identify as $b_j$ and $b_{j+1}$. The only
remaining thing to prove is that there is a zero in the interval
$]-\infty,a_1[$ when $f_1$ is negative. In this case, we use the fact that
$\sum f_k$ is positive, so that the dominant term near $-\infty$ is negative.
The potential is positive at $0-$ hence there must be  a zero in between. Since
we identified $N-1$ zeroes, we have exhausted the whole set of zeroes for the
potential and there cannot be any other, proving in particular that all zeroes
are real.

\subsection{The quantization conditions.}

In fact, using any of the two WKB approaches, we get the same semiclassical
quantization conditions:
\begin{equation}\label{semi}
	\int_{b_j}^{a_{j+1}} dt\;\sqrt{-{v\over4}{\prod (t-b_i) \over \prod
(t-a_k)}} = (n_j  + {\textstyle 1\over4})\pi \hbar
\end{equation} 
in the case where all the monodromies are one and $b_j$ is in the interval
$[a_j,a_{j+1}]$. The integer $n_j$ is the number of zeroes of the wave
function in the interval $[a_j,a_{j+1}]$. If the monodromy around $a_{j+1}$ is
$-1$, we should add one half to $n_j$. In the cases where $b_j$ is near
$a_j$ or becomes smaller, the $\pi/4$ phase factor is no more a good
approximation and we will postpone to further studies the relevant
phase factors. This equation has already appeared in the literature, for
example in~\cite{BaTa92,Gurarie}, where the refinement with the $\pi/4$ phase
first appears in the last paper. It is introduced there using the machinery of
Maslov indices. In this case, the justification through the connection formulae
is very direct, it simply uses the linear approximation of the potential around
the turning point and consequently the asymptotic analysis of the Airy
function:$$
Ai(x) \sim_{+\infty} {1\over 2\sqrt\pi x^{1/4} } e^{-{2\over3}x^{3/2}},
\qquad Ai(-x) \sim_{+\infty} {1\over \sqrt\pi x^{1/4} } \cos( {2\over3}x^{3/2}
- {\pi\over4})
$$
\begin{figure}[t]
\includegraphics[angle=270,width=15cm]{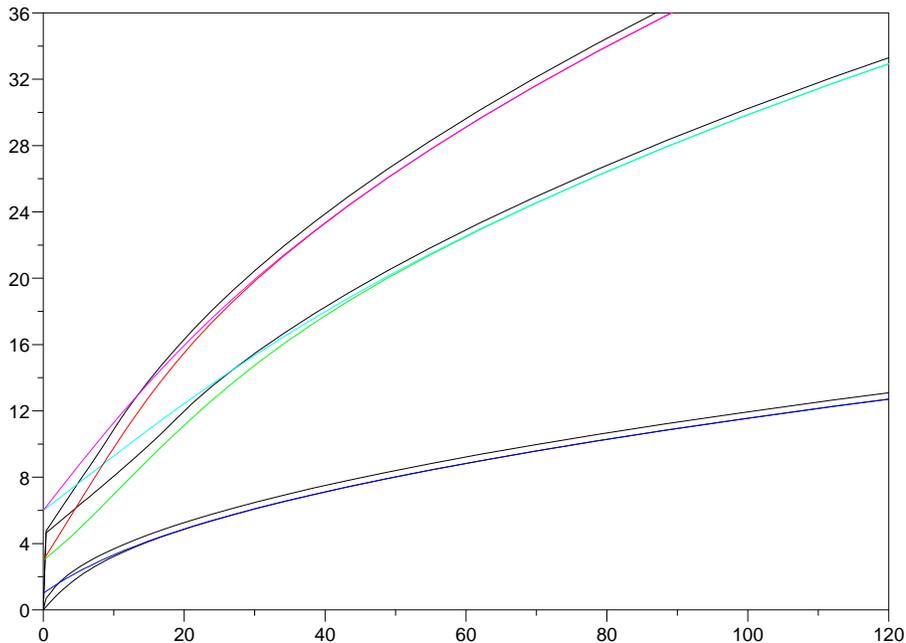}
\caption{Energy levels for $y=2$.\label{ener1}}
\end{figure}
\begin{figure}[t]
\includegraphics[angle=270,width=15cm]{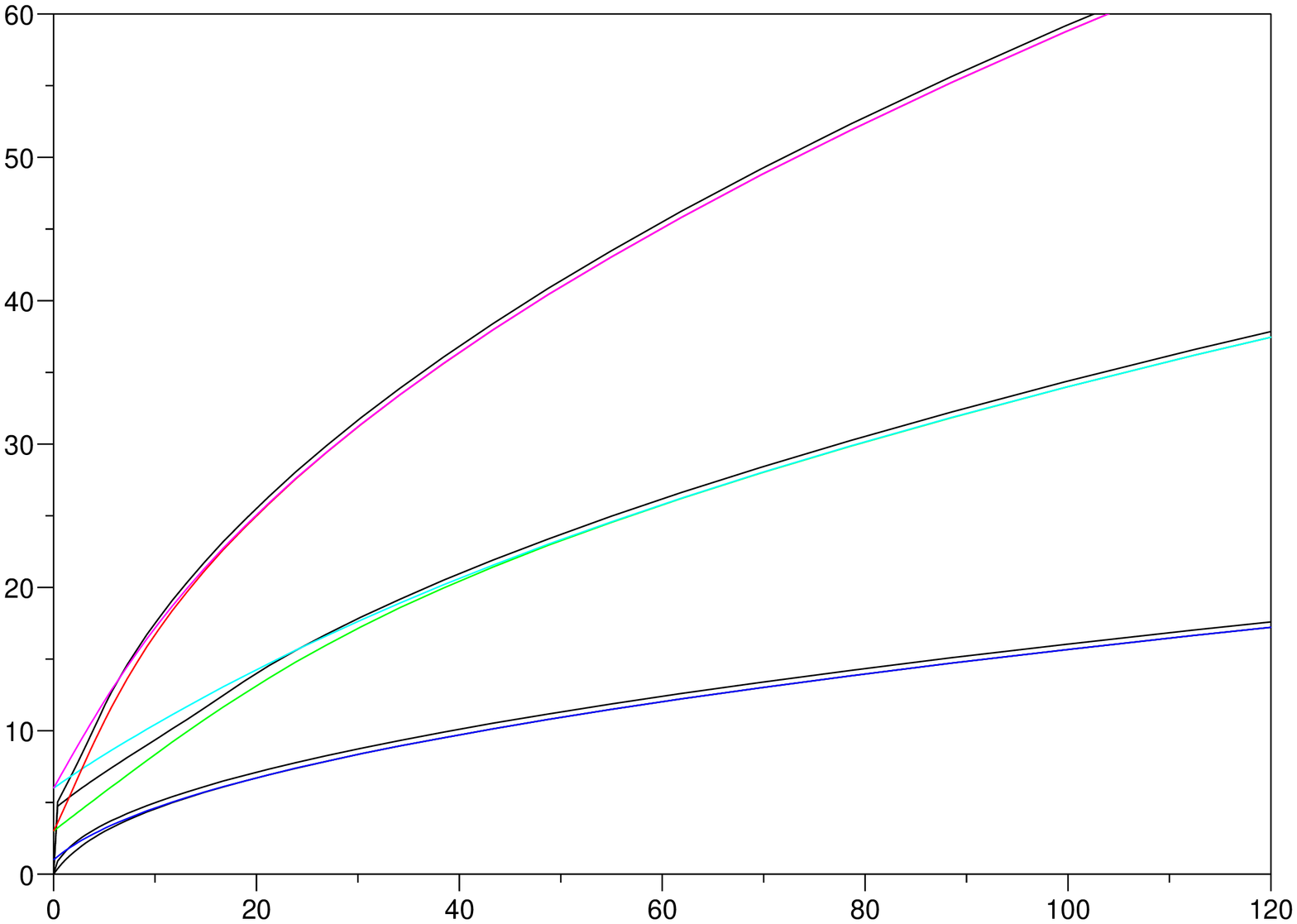}
\caption{Energy levels for $y=5$.\label{ener2}}
\end{figure}
In order to compare the semiclassical predictions to the exact numerical
solutions found in~\cite{BT05}, we solve numerically the quantization
conditions~(\ref{semi}) and plot the resulting expression for the energy as a
function of $v$ compared with two exact energy levels which are degenerate in
the large $v$ limit. We do this for a small number of energy levels in order to
keep the graphs uncluttered and plot them for oscillator strengths equal to
$(0,1,2)$ in Fig.~2 and $(0,1,5)$ in Fig.~2, both in the
case $N=3$. In either figure, the three heavy lines are the WKB predictions. We
see that the small $v$ regime is poorly accounted for, which is of no surprise
since in this case, we are no longer in the situation A) of Fig.~1 and their
should be phase corrections different from $\pi/4$ and moreover, we are in the
deep quantum regime, where we do not expect WKB approximation to be good.
Nevertheless, we plan to study in further works the different cases of Fig.~1
and the transitions between them.

The most notable feature of the curves plotted in Fig. 2 and~3 is that in the
large $v$ limit, there is a small constant separation between the exact and WKB
curves. This separation is independent both on the energy level and the
oscillator strengths. This contradicts the naive expectation that the
corrections to the WKB approximation are of order $\hbar^2$ with respect to the
order zero solution which has an energy of order $1/\hbar$. This order of the
correction is true for the solution away from the turning points, 
but the difference from the Airy function may create
order $\hbar$ correction to the phase as noted in~\cite{Lan38}. It is however
remarkable that this finite correction to the energy does not depend on the
level, so that the transition energies have vanishing errors in the small
$\hbar$ regime. It is the purpose of the following section to show through
perturbative computations the validity of this result.

\section{Large $v$ study.}

We first compute perturbatively for large $v$ or equivalently small
$\hbar$ the energy levels as  obtained from the ``enhanced''
Bohr--Sommerfeld rules, and then as perturbations of quantum oscillators
sitting around the pole of the sphere.

\subsection{Semi--classical calculation.}

We limit ourselves to the $N=3$ case and we fix the oscillator strengths to
$(0,1,y)$ as in~\cite{BT05}. Since $v$ is large, the point on the sphere is
limited to the vicinity of the points $(\pm1,0,0)$, whose
separated variables are $t_1=1$ and $t_2=y$. With $b_1$ close to 1 and $b_2$
close to $y$, $t_1$ is classically limited to the interval $[b_1,1]$ and $t_2$
to the interval $[b_2,y]$, which limits the point to the vicinity of the pole.
To do our perturbative calculations, we will write $b_1=1-\alpha$ and $b_2 = y
- \beta$, so that $\alpha$ and $\beta$ are small. The integrand in the
Bohr--Sommerfeld rules takes the form:
\begin{equation}
\sqrt {-{\frac { \left( t-1+\alpha \right)  \left( t-y+\beta \right) }
{t \left( t-1 \right)  \left( t-y \right) }}}
\end{equation} 
When integrating in $[b_1,1]$, we cannot expand the factors $t-1+\alpha$ and
$t-1$, but all the others can be expanded around $t=1$ and in powers of $\beta$.
The resulting integrals are easily computable yielding:
$${1\over2}\,\alpha\,\pi +{1\over16}\,{\alpha}^{2}\pi -{1\over4}\,{\frac
{\alpha\,\beta }{y-1}}\,\pi + \cdots $$
Similarly for the integral in $[b_2,y]$ we keep the factors $t-y+\beta$ and
$t-y$ and expand the others around $t=y$ in powers of $\alpha$. We then get the
integral:
$$
{1\over2}\,{\frac{\beta }{\sqrt {y}}}\,\pi+
{1\over16}\,{\frac {{\beta}^{2}}{{y}^{3/2}}}\,\pi +
{1\over4}\,{\frac {\alpha\,\beta}{\sqrt {y} ( y-1 )}}\,\pi + \cdots $$
We write that the first integral is equal to
${\hbar\pi\over\sqrt{v}}(n_1+{1\over2})$ and similarly the second equal to
${\hbar\pi\over\sqrt{v}}(n_2+{1\over2})$ which determines $\alpha$ and $\beta$:
\begin{eqnarray*}
\alpha&=&  2\hbar(n_1+{1\over2})\left( {1\over\sqrt{v}} -{\hbar\over v}
\;\Bigl({1\over 4}(n_1+{1\over2})
-(n_2+{1\over2}){\sqrt {y} \over (y-1)}\Bigr)+ \cdots \right) \\
\beta&=&2\hbar (n_2+{1\over2})\left( \sqrt{y\over v} -{\hbar\over v}\;
\Bigl({1 \over 4}(n_2+{1\over2})+(n_1+{1\over2}){\sqrt {y} \over (y-1)}
\Bigr)+ \cdots \right)
\end{eqnarray*}
Note that this justifies a posteriori the claim we made that $\alpha$
and $\beta$ are small, and become smaller when $v$ increases.
Finally the energy $E={1\over2}v(1+y-b_1-b_2)={1\over2}v(\alpha+\beta)$ takes
the simple form, in
which there is no mixing between $n_1$ and $n_2$ excitations:
\begin{equation}
\label{niveausc}
E_{WKB}=\hbar\sqrt{v}\left((n_1+{1\over2})+\sqrt {y}(n_2+{1\over2})\right)
-{\hbar^2(n_1+{1\over2})^2\over 4}-{\hbar^2(n_2+{1\over2})^2\over 4}
\end{equation}
Of course the first term corresponds to the energy of two independent harmonic
oscillators of respective excitations $n_1$ and $n_2$, and nicely fits
the $\sqrt{v}$ look of the energy levels at large $v$  apparent in the figures
of~\cite{BT05}. The second and third term shift these square root terms towards
their correct position, but miss it by a small constant.  As we show below this
small constant is accessible to a more correct quantum computation.

\subsection{Quantum calculation.}

The principle of this calculation is to work out a perturbation of harmonic
oscillators. We detail it in the case $N=3$ and it is then convenient to
go to the elliptic parametrization given by eq.~(\ref{ellparam}).  We
expand the change of variable in the vicinity of $t=1$ and
$t=y$. Taking the origin in $t=y$, we write 
\[x= \int { d\tau \over \sqrt\tau} {1\over\sqrt{y(y-1)}}\Bigl( 1 + {\tau\over 2
y} + {\tau \over 2(y-1)} + \cdots\Bigr) \]
which can be inverted to yield
\begin{equation}
 \tau = y-t = {1\over4}y(y-1) x^2 - {1\over48} y (y-1) (2y-1) x^4 + \cdots
\end{equation} 
A similar computation gives the parametrization around $t=1$
\begin{equation}
 \tau = 1-t = {1\over4}(y-1) x^2 - {1\over48} (y-1) (y-2) x^4 + \cdots
\end{equation} 
These values of the variable $t$ will be used in the
Schr\"odinger equation~(\ref{ellequ}) to express in terms of the variable $x$
the potential $\pm(v/4\hbar^2)(t-b_1)(t-b_2)$. The sign can change since in the
calculation of $x$, we forgot the imaginary unit $i$ which can appear according
to the sign of the term under the square root.

In the large $v$ limit, $b_1$ is near $1$ and $b_2$ is near $y$ and the lowest
relevant order for the potential for $t$ around $y$ is:
$$ {1\over4\hbar^2} v(y-1)\Bigl( (y-b_2) - {1\over4}y(y-1) x^2 \Bigr) $$
which is the potential for a harmonic oscillator with energy
$(v/4\hbar^2)(y-1)(y-b_2)$ and oscillator strength $(v/4\hbar^2)y(y-1)^2$.
This gives $y-b_2= 2\hbar\sqrt{y/v} ( n_2+ 1/2)$, which is exactly the
semiclassical result at this order. Similarly, $1-b_1=2\hbar/\sqrt{v}
(n_1+1/2)$. In order to choose the terms which will contribute to the next
order, we should remember that for the harmonic oscillator, $x^2$ is of order
$v^{-1/2}$. We therefore get corrections stemming from the variation of the
oscillator strength, from the factor in front of $y-b_2$ and from 
anharmonic terms in $x^4$ which stem in part from the one in $\tau$ and in part
from the product of the two $x^2$ terms occurring in both factors $t-b_i$.
The order one corrected potential is:
\begin{eqnarray*}
{1\over4\hbar^2} v  \mkern -20mu &&\left(
(y-b_2)\bigl(y-1+{2\hbar\over \sqrt{v}}(n_1+{1\over2})\bigr) -{1\over4}y(y-1)^2
x^2\right. \\
&&-{\hbar\over2\sqrt{v}} \Bigl( n_1+{1\over2}+
(n_2+{1\over2}) \sqrt y \Bigr)y(y-1)  x^2 \left.
+{1\over48}y(y-1)^2(5y-1)x^4\right)\\
\end{eqnarray*} 
With the anharmonic oscillator Schr\"odinger equation written as $ - \Psi''+1/4
(x^2+\lambda x^4)\Psi = E \Psi$, first order perturbation theory yields the
energy levels $E_n = n + 1/2 +3/4 \lambda (2n^2+2n+1)$. This result is easily
obtained writing $x^4$ in terms of creation and annihilation operators.
Putting these corrections together, we get:
$$y-b_2=2\hbar \sqrt{ y\over v} ( n_2+{1\over 2})-
{\hbar^2\over v} \Bigl({2\sqrt{y}\over (y-1)}(n_1+{1\over 2})(n_2+{1\over 2}) +
{1\over 2}(n_2+{1\over 2})^2 +{1\over 8}{5y-1\over y-1}\Bigr)$$
Note that this only differs from the semiclassical result by the last term,
independent on the $n_i$.

A similar computation yields the potential developed around $t=1$ and the first
order correction to the energy of the corresponding anharmonic oscillator gives
$$ 1-b_1 = 2\hbar { 1\over \sqrt v} ( n_1+{1\over 2})+
{\hbar^2\over v} \Bigl({2\sqrt{y}\over (y-1)}(n_1+{1\over 2})(n_2+{1\over 2}) -
{1\over 2}(n_1+{1\over 2})^2 -{1\over 8}{y-5\over y-1}\Bigr)$$
Whatever happens outside of these two regions is exponentially suppressed for
large $v$, since the wave function of the harmonic oscillator are
already exponentially small there. Combining the values of $b_1$ and $b_2$, we
obtain for the energy at this order in $1/v$ a result which differs from the semiclassical one of eq.~(\ref{niveausc}) only by
a constant
$$ E_{Q} = E_{WKB} - {3\over 8} \hbar^2 $$
This nicely reflects our findings on Figs.\ 2 and~3 where $\hbar = 1$.

\section{Conclusion.}
The interaction of the study of the Neumann model and the generalized Lam\'e
equation allowed us to give precise results on the behaviour of the spectrum in
some limits. The generalized Lam\'e equation has independent interest since it
is the most general linear differential equation of second order with only
regular singularities at finite distance. 

We were able to precisely count the different cases for the positions
of the zeroes of the potential, showing in particular that they are all real.
It is also remarkable that there is a constant error in the WKB approximation.
This error is however independent on the level, so that it does not appear in
transition energies. In this integrable case, the WKB approximation gives pretty
good results even for low excitation numbers and the further study of its
behaviour around turning points at low energy could make it even more useful.

\end{document}